\documentclass{sf2a-conf2016}
\usepackage{graphicx}
\usepackage{hyperref}
\usepackage[]{natbib}  
\usepackage{epstopdf}
\usepackage{amssymb}
\usepackage{amsmath}

\def\mean#1{\left< #1 \right>}

\def\BibTeX{{\rm B\kern-.05em{\sc i\kern-.025em b}\kern-.08em
    T\kern-.1667em\lower.7ex\hbox{E}\kern-.125emX}}
\bibpunct{(}{)}{;}{a}{}{,}  

\begin{document}

\TitreGlobal{SF2A 2016}


\title{Review of gas and dust in debris discs}

\runningtitle{Debris disc review}

\author{Quentin Kral}\address{Institute of Astronomy, University of Cambridge, Madingley Road, Cambridge CB3 0HA, UK}





\setcounter{page}{237}


\maketitle


\begin{abstract}
This proceeding summarises a talk given on the state-of-the-art of debris disc modelling. We first review the basics of debris disc physics, which is followed by a short overview of the state-of-the-art in terms of modelling dust and gas in debris disc systems.
\end{abstract}



\section{Introduction}
Circumstellar discs are observed around an increasing number of stars. In this proceeding, we are interested in debris discs rather than younger protoplanetary discs still full of gas and where giant planets are still forming. These debris discs are left-overs of planetary
formation, old, cold, gas-poor and are detected owing to their infrared (IR) excesses. These discs contain solid bodies with size ranging from a few 100km to a few microns. The mass reservoir is located in the biggest bodies that collide, fragment and create smaller planetesimals. 
These smaller fragments collide and are ground down to dust.
This large cross section of dust or debris can then be observed from Earth, either in thermal emission in the far-infrared or in scattered light in the optical or near-IR. The Kuiper and asteroid belts are the debris disc of our Solar System. However, our debris disc is critically different from
 observed debris discs; it is less massive and less collisionally active and thus would be impossible to detect with current instruments.

\begin{figure}[ht!]
 \centering
 \includegraphics[width=0.29\textwidth,clip]{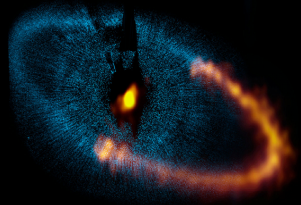}%
\includegraphics[width=0.21\textwidth,clip]{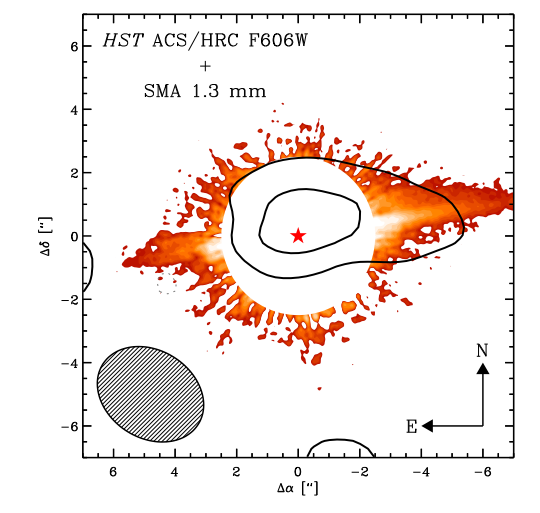} 
\includegraphics[width=0.25\textwidth,clip]{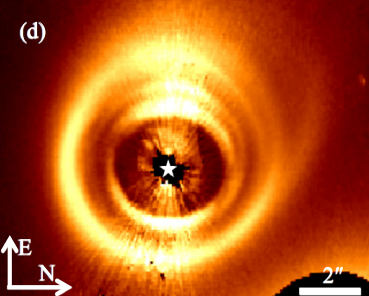}           
  \caption{{\bf Left:} Fomalhaut \citep{2012ApJ...750L..21B,2013ApJ...775...56K} {\bf Centre:} HD 15115 \citep{2015ApJ...801...59M} {\bf Right:} HD 141569 \citep{2016ApJ...818L..23K}}
  \label{kral:fig8}
\end{figure}

We observe thousands of such debris discs in extrasolar systems. We estimate that at least 25\% of stars show an IR-excess (and therefore possess a debris disc), 
but this number could be much higher if we were able to detect very tenuous Kuiper belt-like systems \citep[e.g.][]{2014prpl.conf..521M}. For the closest and/or the brightest debris discs, we are able to resolve them. As of today, $\sim$ 90 debris discs are resolved at different wavelengths.
On the resulting images, we observe that debris discs are not simple symmetric circular discs but rather, they show many structures such as warps, clumps, spirals, depleted zones, shifts of the centre of the disc compared to the star, asymmetric needle-like shapes, ... 
(see Fig.~\ref{kral:fig8}). These structures inform us about the interactions between the disc and its environment. The main question that now drives debris disc science is to try to understand where these structures originate from.
To do so, complex models are being developed and I will summarise here the state-of-the-art in terms of modelling debris discs.

Over the past few years, an increasing number of gaseous species (atoms and molecules) have been detected around debris discs (in more than 10 systems). The gas is at low levels, critically different from protoplanetary discs. However, it is surprising as debris discs were thought to be gas-free 
after the protoplanetary disc phase and subsequent dispersal of the primordial gas via accretion and photoevaporation \citep[e.g.][]{1981ARA&A..19..137P,2001MNRAS.328..485C}. However, many progresses have been made to understand
where the gas may originate from and to model its thermodynamical evolution that might solve the conundrum.

I will first explain in more details the physics of debris discs and how complicated it can be to model numerically (section~\ref{ddp}). I will
then focus on dust modelling (section \ref{dust}) before ending on the new hot topic in our field focusing on understanding gas observations in debris discs (section \ref{gas}).

\section{Debris disc physics}\label{ddp}

We here summarise the most important forces and effects that, in an ideal world, should be taken into account when modelling debris discs. We will find out in section \ref{dust} that due to a lack of CPU-power, some concessions have to be made concerning which effects are most important
to include in our codes. 

The equation of motion describing the evolution of a solid body of mass $m$ located at a distance $r$ from the central star of mass $M_\star$ is given by

\begin{equation}\label{eqdyn}
  m\frac{{\rm d}^2\vec{r}}{{\rm d}t^2}=-\frac{GmM_\star}{r^2}\vec{e_r}+\vec{F}_{\rm rad}+\vec{F}_{\rm drag}+\vec{F}_{\rm others},
\end{equation}

\noindent where the first term of the right-hand side of the equation is the gravitational force from the central star $\vec{F}_g$, $\vec{e_r}$ is the radial unit vector, $G$ is the gravitational constant, $\vec{F}_{\rm rad}$ is the force due to radiation pressure from the star, 
$\vec{F}_{\rm drag}$ are for all the drag forces and $\vec{F}_{\rm others}$ are other forces that can be exerted on this mass $m$
in a debris disc environment. In Eq.~\ref{eqdyn}, the collisions are not taken into account. Subsection \ref{col} explains what can be their effect on the solid-body dynamics. We now describe each term of the equation of motion.

\subsection{Stellar gravity and radiation pressure}\label{radp}

\begin{figure}[ht!]
 \centering
 \includegraphics[width=0.6\textwidth,clip]{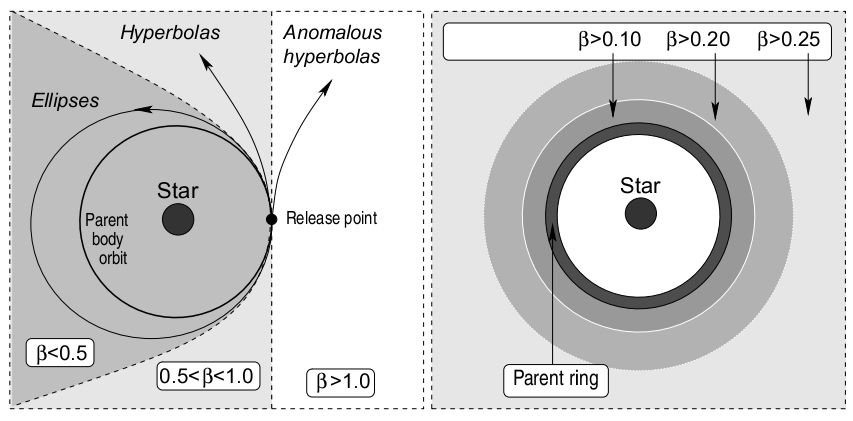}      
  \caption{Different types of orbits in a debris disc resulting from different values of $\beta$ (see subsection \ref{radp}). From \citet{2010RAA....10..383K}.}
  \label{kral:fig2}
\end{figure}

The photons coming from the central star transfer angular momentum to the orbiting dust grains. Only the smallest dust grains are affected by this purely radial force \citep{1979Icar...40....1B} and this translates as the overall gravitational force becoming

\begin{equation}
  \vec{F}_g+\vec{F}_{\rm rad}=-\frac{GmM_\star (1-\beta)}{r^2}\vec{e_r},
\end{equation}

\noindent where $\beta=|\vec{F}_{\rm rad}/\vec{F}_{g}|$. When $\beta>0$, it is as if the grains felt a star with a smaller mass $M_\star (1-\beta)$. $\beta$ is usually modelled as being $\propto 1/s$, the grain size. In reality, it can be more complicated as there are other
dependencies, more precisely

\begin{equation}
 \beta=\frac{3 L_\star \mean{Q_{\rm rad}}}{16 \pi G M_\star c \rho s},
\end{equation}

\noindent where c is the light speed, $\rho$ the bulk density of grains, $L_\star$ the star's luminosity and $\mean{Q_{\rm rad}}$ the mean radiation pressure coefficient (averaged over wavelengths). $Q_{\rm rad}$ depends on the grain compositions and can vary steeply
for one composition to another, or as a function of $s$. One can show that if a grain with $\beta > 0.5$ is created from a parent body on a circular orbit at a distance $a_0$, it will become unbound and will be blown out on a hyperbolic orbit. If  $\beta < 0.5$, the dust grain will
have an eccentric orbit and its new semi-major axis and eccentricity are given by $a=a_0\frac{1-\beta}{1-2\beta}$ and $e=\frac{\beta}{1-\beta}$. These different orbits are well described in Fig.~\ref{kral:fig2}. Therefore, there will be a radial segregation of sizes as the smallest grains with the 
largest eccentricities will reach farther out in the system whilst bigger bodies that are less pushed by radiation pressure will stay close to the parent belt where they are produced.

We note here that we do not take into account any gravitational disturbances that could be created by a planet located close to the disc. The effects of planets are discussed further in subsection \ref{other}.

\subsection{Drag forces}
\begin{figure}[ht!]
 \centering
 \includegraphics[width=0.5\textwidth,clip]{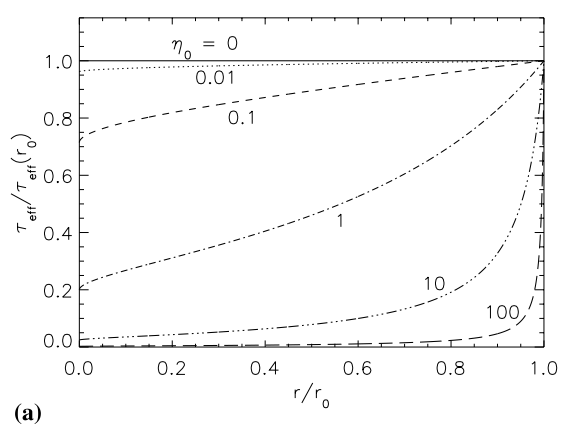}      
  \caption{Optical depth as a function of distance to star at steady-state for a narrow parent belt (located at $r_0$) evolving under PR-drag+collisions as a function of $\eta_0$ (which parameterises the dynamical excitation of the belt, see main text). From \citet{2005A&A...433.1007W}.}
  \label{kral:fig3}
\end{figure}

The radiation pressure force has also an orthoradial component described by the Poynting-Robertson effect. This tangential force is dissipative and opposes the motion, one can show that this force is \citep{1979Icar...40....1B}

\begin{equation}\label{prdrag}
  \vec{F}_{\rm PR}=-\beta \frac{G m M_\star}{r^2} \left( \frac{v_r}{c} \vec{e_r} + \frac{\vec{v}}{c} \right),
\end{equation}

\noindent where $\vec{v}$ is the velocity vector of the dust grain and $v_r$ its radial component. Under this force alone (no collisions), a dust grain spirals towards the central star on a 
timescale equal to $t_{\rm PR}=400\beta^{-1}(r/a_\oplus)^2(M/M_\odot)^{-1}$, where $a_\oplus$ is the semi-major axis of the Earth \citep{2008ARA&A..46..339W}. If collisions were
taken into account the equilibrium would be changed as a particle migrating in by PR-drag could get destroyed by collisions much before reaching the central star. Fig.~\ref{kral:fig3} shows the radial variation of the optical depth for different
dynamical excitations (parameterized by $\eta_0$) of a narrow belt at $r_0$. To do so, \citet{2005A&A...433.1007W} uses a simple analytical model  where all the grains have the same size. The parameter $\eta_0$ used in Fig.~\ref{kral:fig3} is proportional to the optical depth in the parent belt and describe
the dynamical excitation of the belt. When $\eta_0$ is small, collisions are unimportant and the surface density of the transport-dominated tenuous belt is constant with distance to star. When $\eta_0$
is large, the system is collision-dominated and the dust is confined to a narrow belt as it gets destroyed before reaching further in. Fig.~\ref{kral:fig3} also shows intermediate cases for the evolution of 
the optical depth with radial distance to star. More complicated models can get rid of the same size assumption and work out the profile at steady-state or as a function of time \citep[e.g.][]{2011CeMDA.111....1W, 2008ApJ...673.1123L, 2013A&A...558A.121K}.

Around late-type stars where the radiation pressure is smaller, stellar wind particles can matter. Similarly to the radiation pressure forces, stellar wind forces can be decomposed into a radial and tangential force. Eq.~\ref{prdrag} also applies, replacing $c$ by $v_s$, the stellar wind velocity. 
Stellar wind drag can then become important as $v_s \ll c$ \citep[e.g.][]{2011A&A...527A..57R}.

Finally, when a dust grain orbits in a gas disc, it feels a gas drag that can be modelled by the Epstein force \citep{2001ApJ...557..990T}. The gas drag timescale is given by $\rho s/ (\rho_g v_{\rm th})$, where $\rho_g$ is the gas number density, $\rho$ the bulk density of grains
and $v_{\rm th}$ is the thermal velocity. Around the gas-rich system $\beta$ Pic this force do not affect bound grains \citep{2016MNRAS.461..845K} but for more gaseous systems that are in transition to becoming debris discs and still contain 
a fair amount of primordial gas, it may well be that dust is affected by gas drag.

\subsection{Collisions}\label{col}
In debris discs, mass flows from the biggest solid bodies ($\sim$ 100km) to the smallest grains (micron size) owing to collisions: this is called the collisional cascade and always replenishes the smallest grains that are blown out from the system as soon as they are small
enough to be blown out by radiation pressure (i.e. $\beta \gtrsim 0.5$). Collisions happen at high-velocity in debris discs (a few km/s), which often lead to fragmentation of the colliding bodies. The biggest bodies from the mass reservoir can collisionally survive billions of years \citep{2008ApJ...673.1123L} before
being ground down to dust. An important quantity to work out the outcome of a collision is the critical specific collisional energy

\begin{equation}
 Q_D^*=Q_s \left ( \frac{s}{1 {\rm m}} \right)^{-b_s}+Q_g \left ( \frac{s}{1 {\rm km}} \right)^{-b_g},
\end{equation}

\noindent which is often characterised as two power laws and is the impact energy per unit mass of the target that results in a biggest fragment that has half the mass of the initial target. $Q_D^*$ varies with the impactor size. Indeed, lab experiments (with small grain collisions) 
and numerical simulations (for large body collisions) show that smallest grains become more coherent and thus harder to destroy and larger bodies can re-accrete fragments owing to their larger gravity (making them harder to destroy). Depending on the impact energy,
a given collision can end up in a fragmentation, craterisation or even just a rebound (but this is rarely the case in debris discs). The physics of collisions is complex as it covers 12 orders of magnitude in size and can result in several outcomes
depending on the respective sizes of the target and impactor and their relative velocity. Analytical solutions show that the resulting particle size distribution at steady-state scales as $s^{-3.5}$. More refined numerical models show that it is more complex and
4 different regimes exist: the small eccentric grain regime, the strength regime, the gravity regime and lastly the primordial regime \citep[e.g.][]{2007A&A...472..169T}.

\subsection{Other effects}\label{other}
Some other effects can be at play in debris disc environments. For instance, the local flux of neutral atoms in the interstellar medium ($\sim$ 0.1 cm$^{-3}$) can affect the dynamics of dust grains in debris discs that are not too collisionally active \citep[i.e when the optical 
depth $\lesssim 10^{-5}$,][]{2011MNRAS.416.1890M}. This effect can
create some potentially observable butterfly-like shapes \citep[e.g. the debris disc around HD 61005][]{2016A&A...591A.108O}.
Recently, magnetic fields were proposed to be a potential mechanism to trap dust that would drift very close to the star \citep{2016ApJ...816...50R,2016ApJ...818...45S}. Sublimation can also affect the dynamics of grains when a particle drifts inwards by PR-drag.
Indeed, grains when sufficiently heated sublimate and become smaller. This induces radiation pressure to be more effective to push grains, which can potentially be sent out again \citep[e.g.][]{2009Icar..201..395K}.
Also, planets in a debris disc system can influence the dynamics of solid bodies. Resonances can trap dust and create clumps \citep[e.g.][]{2003ApJ...598.1321W}. Long term secular effects from an eccentric planet can induce an eccentricity
to the disc and offset its centre compared to the star \citep[the so-called pericentre glow,][]{1999ApJ...527..918W}. Also, a planet really close to the disc can truncate its inner edge and steepen its surface density profile because of the clearing of its chaotic zone
 \citep[e.g. the narrow disc around HR 4796,][]{2012A&A...546A..38L}.




\section{Dust modelling}\label{dust}
We note that the whole field cannot be summarised in this brief proceeding and we will focus on the most state-of-the-art modelling methods that have been used recently. 

\subsection{Radiative transfer}
Dust radiative transfer models were originally used to predict the best-fit dust distribution from a system's SED. However, this is a degenerate process as a bigger grain closer-in will produce the same flux as a smaller grain further out.
These radiative transfer codes are now refined and fit at the same time the SED with resolved images and interferometric nulls, which break most of the degeneracies. 

One recent example is the study of $\eta$ Corvi by \citet{2016ApJ...817..165L}. They use the radiative transfer code GRaTeR \citep{1999A&A...348..557A} to fit both the SED, Herschel images and KIN nulls against a large parameter space of different discs.
Doing so, they were able to constrain the position of the outer belt and show marginal evidence for asymmetries, but also, they could determine the spatial distribution of the inner belt (exozodis) at a sub-au scale. They use Bayesian statistical analysis 
and integrate over each parameter of the model to work out a probability density for each disc and derive uncertainties. They were able to probe
the best-fit compositions for both discs. The best-fit model can then be used to make predictions for observability with other instruments (ALMA, JWST, ...) and predict what more could be assessed through these new observations. 

Another state-of-the-art study showing the capacity of radiative transfer modelling is Milli et al. (2016, recommended for publication). From SPHERE very high-angular resolution images of HR 4796A, they were able to derive the scattering phase
function (SPF) of the dust up to very small angles (13.6$^\circ$). It shows a peak of forward scattering for scattering angles below 30$^\circ$ and confirm the side of the disc inclined towards the Earth. Using the Mie theory assuming
spherical dielectric grains, they computed SPFs for a large range of compositions, porosity, size distributions, ... Thanks to this modelling (and performing a Bayesian analysis), they predict that the dust population is dominated by large particles ($\sim$ 20$\mu$m)
well above the size that usually dominates the cross section, i.e close to the blow-out size. The SPF also gives some constraints on the size distribution. One should then try to understand this prediction using one of the models described in the next subsections. 

\citet{2016A&A...592A..39K} used radiative transfer to check the impact of disc asymmetries on astrometric measurements. They find that new missions such as Theia \citep[new name for NEAT,][]{2012ExA....34..385M} that aim to detect small Earth-like planets with high precision astrometry could be affected
by dust clouds (of cometary mass) that would create an astrometric signal of the same order of magnitude as an Earth-like planet. Moreover, current infrared missions could not detect IR-excesses of such small clouds that would create such a fictitious Earth-like astrometric signal. They find
that there are ways to disentangle a dust cloud from a real planet (e.g. by observing at different wavelengths), and it also means that it will be a new way to discover asymmetric close-in discs that cannot be resolved or detected otherwise. 

\subsection{Collisional approach}
Collisional modelling has been used in our community to understand the origin of the observed IR-excesses. Thanks to this approach, we could assess the mass reservoir needed as well as the planetesimal eccentricities required to create the right amount of dust and follow the time evolution
of the planetesimal grinding process. These codes (using the particle-in-a-box approach) have now reached a high level of sophistication \citep[e.g.][]{2007A&A...472..169T} but still treat the dynamics poorly (e.g. no azimuthal dependence). We here give recent examples of what can be achieved with this approach.

\citet{2015A&A...581A..97S} studied the AU Mic (M-dwarf) debris disc with their sophisticated collisional code called ACE \citep{2006A&A...455..509K,2008ApJ...673.1123L}. They wanted to find the best disc model that would reproduce both the SED, the ALMA image (1.3mm) as well as scattered and polarised light data.
Running their model they can find the best radial and size distributions of the particles in the whole disc and constrain the position of the outer edge of the belt as well as the preferred dynamical excitation in the belt ($e<0.03$). 
They also find that the stellar mass loss rate should exceed the solar one by a factor $\sim$50 (which is expected around M stars) for the stellar wind to be able to drag enough material in. 

Recently, \citet{2014A&A...566L...2K} pointed out that the minimum fragment size that can be produced after a collision is limited by the conservation of energy. \citet{2016A&A...587A..88T} implemented this new constraint in his code to study the observational effects it might have. It was thought
that it could explain the mysterious fact that we observe grains that are much larger than the blow-out size \citep{2015MNRAS.454.3207P}. The use of this collisional code could prove that the minimum fragment size dependence has a weak effect on the predicted size distribution. Therefore, the intriguing result
by \citet{2015MNRAS.454.3207P} has not yet been fully explained by debris disc models.

\subsection{Purely dynamical approach}
When one tries to reproduce complex structures observed in resolved images of debris discs, the N-body approach is widely used. It allows to take into account complex interactions with planets or a companion star but the collisions are 
totally neglected. It can still be useful to study purely dynamical effects on the biggest bodies and derive some general results that may be altered by collisions. We will now describe two new papers that used that type of approach recently, in conjunction with ALMA images.

The traditional thought that gaps in debris discs are created by a planet within the gap has been revisited recently. \citet{2015MNRAS.453.3329P} showed that a double ring system could be created from an eccentric planet of mass comparable to the disc mass that is located near the inner
edge of the innermost disc and not between the two belts. Indeed, the initially eccentric planet is circularised by interactions with the disc and the secular effect of the planet can cause debris to apsidally antialign with the planet's orbit, clearing a larger region than a higher mass planet would
and, therefore, creating a double belt shape. This scenario could potentially explain the shape of the disc around HD 107146 recently observed with ALMA.

The HR 8799 double belt star was recently imaged with ALMA \citep{2016MNRAS.460L..10B}. This system possesses 4 giant planets that are observed between the two belts. The inner edge of the outer belt could be resolved with ALMA for the first time, and is located at $\sim$ 145au.
The outermost planet called HR 8799b has a semi-major axis $\sim$68au \citep{2016A&A...587A..57Z}, which may be too small to sculpt the inner edge of the outer belt at 145au. Running N-body simulations or using analytical estimates, one can compute the chaotic zone
of planet b (within which mean motion resonances overlap), which depends on the mass and eccentricity of the planet. Using this approach, they find that the planet b chaotic zone could clear objects up to 110au, which opens up for the possibility of having an additional, yet hidden, planet in the system. This
potential planet would be located between 110 and 140au and would be lighter than 1.25M$_{\rm J}$. \citet{2014MNRAS.440.3140G} also ran simulations with the system of 4 (or 5) planets to test for the stability of the planetary system as a whole and see whether it is in a stable configuration. They find that
the planets are most likely in a double Laplace resonance to explain the stability of this packed system over the age of the star ($\sim$160Myr).

\subsection{Refined approach coupling dynamics and collisions}
A big step forward has been made over the past few years, when the first codes coupling collisions and dynamics emerged. Amongst the most sophisticated debris disc models that have
been developed to date are the DyCoSS and LIDT-DD codes \citep[see also][]{2012AJ....144..119L,2013ApJ...777..144N}. Both can follow the collisions at the same time as the dynamics. However, they are different in their principles and limitations. DyCoSS is restricted
to steady-state situations under the influence of a single planet and collisions are fully destructive. LIDT-DD overcomes these limitations but the price to pay is a slower computational time.

\begin{figure}[ht!]
 \centering
 \includegraphics[width=0.5\textwidth,clip]{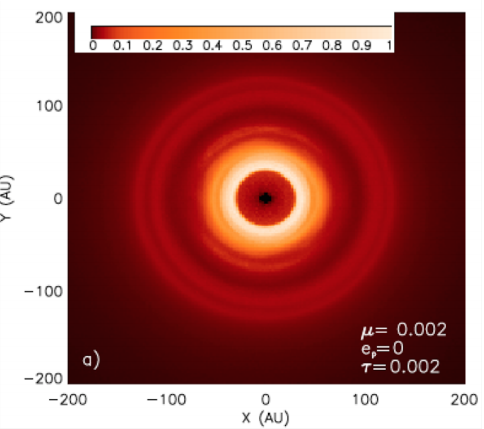}      
  \caption{The effects of a planet at 75au embedded within a broad disc (30-130au) modelled with the partial coupling code called DyCoSS.}
  \label{kral:fig4}
\end{figure}

DyCoSS can study very fine spatial structures and has been used to study
debris discs around binaries \citep{2012A&A...537A..65T} or discs with an embedded or exterior planet \citep{2012A&A...547A..92T,2012A&A...546A..38L}. 
One such simulation with a planet embedded in a broad disc is shown in Fig.~\ref{kral:fig4}. Owing to the presence of the planet,
resonant structures develop as well as a density gap at the planet position. According to previous N-body non-collisional simulations, the density gap, which is the chaotic zone
of the planet should be totally devoid of dust. However, this new generation code is able to show that it is more complicated, as small grains produced in the inner disc that are on
eccentric orbits, actually fill up the chaotic zone.

\begin{figure}[ht!]
 \centering
 \includegraphics[width=0.8\textwidth,clip]{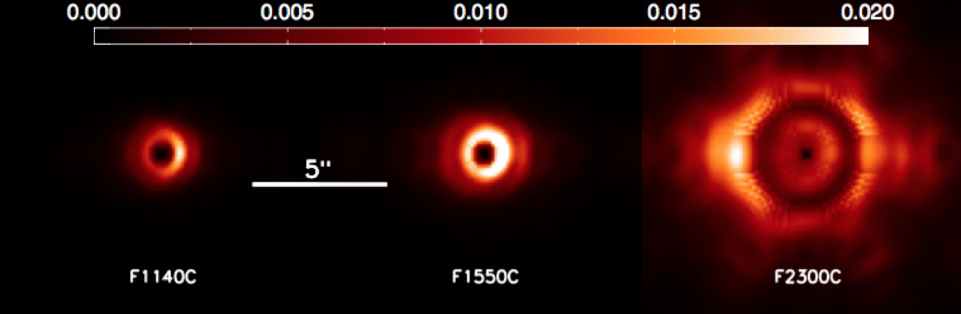}      
  \caption{MIRI/JWST synthetic observation of the aftermath of a massive collision between two asteroid-like objects. The asymmetry on the right side at 11 and 15 microns and on the left side at 23 microns is a typical signature of these events.}
  \label{kral:fig5}
\end{figure}

LIDT-DD \citep[described in][]{2013A&A...558A.121K} is able, for the first time, to treat collisions and dynamics in a self-consistent fashion and allows to follow the time evolution till steady-state of the many fragments that are produced during collisions.
\citet{2015A&A...573A..39K} presents the first astrophysical application of the code, which follows the evolution of violent collisions between sub-planetary mass bodies that are expected to happen in the late stages of planetary formation.
This new generation model is able to tackle such an arduous problem for the first time, and leads to some interesting
results such as providing the brightness of such violent phenomena, their timescale, their detectability, as well as being
able to predict an infallible signature of such events. These giant impacts create a strong brightness asymmetry at the collision point that could be observed with SPHERE (for the closest systems) or with MIRI/JWST in the mid-IR (see Fig.~\ref{kral:fig5}).
Confirmed detections of this signature would lead to actual observations of on-going
planetary formation, which would be a major advance in our understanding of planetary formation.

\section{Gas modelling}\label{gas}

The search for gas around main sequence stars is becoming a hot topic as it is a new way to probe volatiles in planetary systems, where planets have already formed. Molecular (CO) and atomic species (carbon, oxygen, metals)
are now detected around more than 10 main sequence stars \citep[e.g.][]{2004A&A...413..681B,2006Natur.441..724R,2014Sci...343.1490D,2016A&A...591A..27B}. This was not expected as planetary systems were thought to be gas-free after the protoplanetary disc phase and subsequent dispersal of the primordial gas via accretion and photoevaporation 
\citep[e.g.][]{1981ARA&A..19..137P,2001MNRAS.328..485C}.
The majority of the observed molecular gas in these $\sim$ 10 systems is presumably not primordial as the CO photodissociation timescale is on the order of 100 years and primordial CO would be long gone. We suppose that for most cases, the observed gas is secondary. Also, gas 
in these old systems is only observed around stars with a debris disc. We expect that the secondary gas is created from the volatile-rich solid bodies of debris discs, either by photodesorption \citep{2007A&A...475..755G} or solid-body collisions \citep{2007ApJ...660.1541C,2012ApJ...758...77Z}.

\begin{figure}[ht!]
 \centering
 \includegraphics[width=0.5\textwidth,clip]{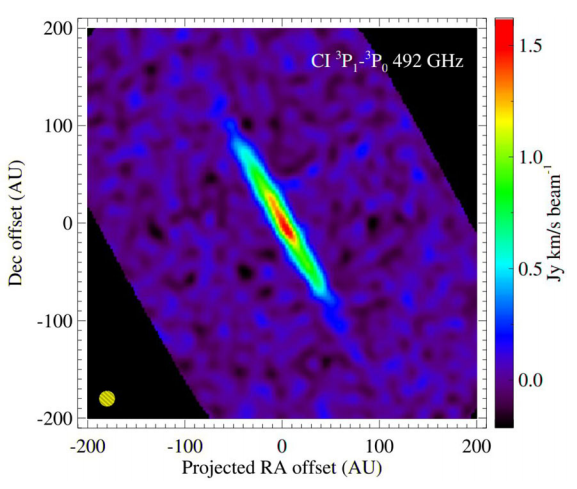}      
  \caption{Predicted ALMA CI map at 610 microns from \citet{2016MNRAS.461..845K} best-fit model.}
  \label{kral:fig6}
\end{figure}

A new model has recently been proposed by \citet{2016MNRAS.461..845K} to describe the thermodynamical evolution of such secondary gas in planetary systems. This model has been used on the famous $\beta$ Pic system, for which the largest number of species have been detected so far \citep[see also][]{2013ApJ...762..114X}.
The model is able to reproduce all $\beta$ Pic gas observations (and make predictions for future observations, see Fig.~\ref{kral:fig6}) and may, more generally, be used to explain the gas origin and dynamics around all debris discs. The model proposes that 1) CO is produced from volatile-rich solid bodies located in debris belts. 
2) CO photodissociates in less than 120 years in C+O. 3) The carbon and oxygen atoms evolve by viscous spreading (parameterised with an $\alpha$ prescription), resulting in an accretion disc inside the parent belt and a decretion disc outside. 
The thermodynamical model follows the dynamical evolution of atoms as well as their ionisation fractions, excitation and temperature. The viscosity may come from the magnetorotational instability that might be active in debris discs as proposed by \citet{2016MNRAS.461.1614K}.

\begin{figure}[ht!]
 \centering
 \includegraphics[width=0.9\textwidth,clip]{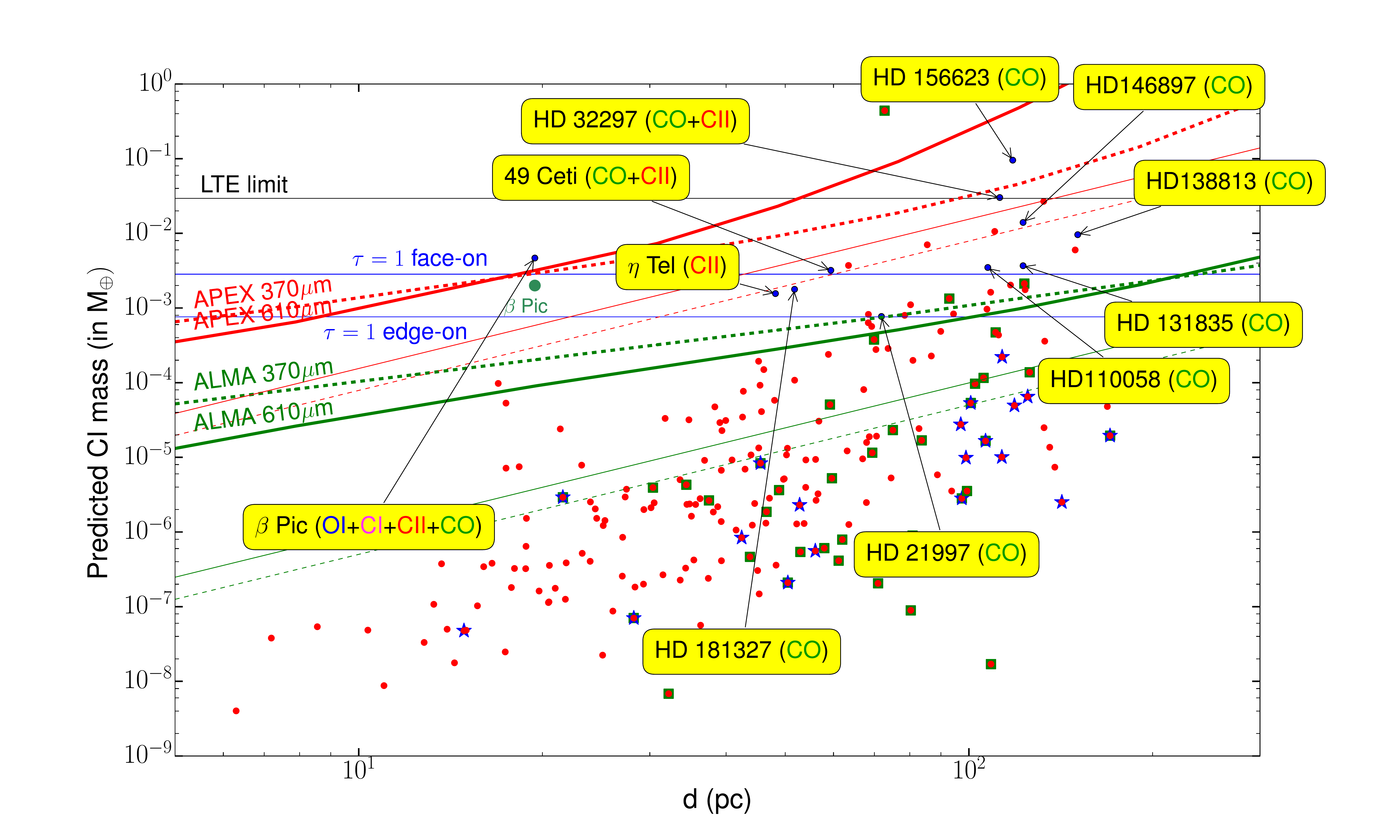}      
  \caption{Predicted CI mass (in $M_\oplus$) as a function of distance to Earth ($d$). Planetary systems with gas detections are annotated using yellow boxes and the elements detected within each system are written in between parenthesis. The CI mass for $\beta$ Pic,
 derived from \citet{2016MNRAS.461..845K} is shown as a green point.
 The red points are predictions for debris disc' systems without gas detected. Detection limits at 5$\sigma$ in one hour are shown for APEX (in red) and ALMA (in green) at 370 (dotted) and 610 microns (solid). The thin lines are for LTE calculations and thick lines
for more realistic NLTE calculations.
The blue lines show the limit above which the CI line gets optically thick ($\tau > 1$) for both edge-on and face-on cases. The black line is the position under which LTE is not a good approximation anymore (assuming a belt located at 85au). 
The blue star and green square symbols show systems that are too warm to still retain gas or have a small photodissociation timescale, respectively.}
  \label{kral:fig7}
\end{figure}

This secondary gas model can be applied to a large sample of debris disc systems. Kral et al. (in prep) assume that the gas production rate depends on the debris disc properties. Indeed, the more collisional and massive is the disc, the more gas is expected to be released. They can
then compare their secondary gas model predictions to existing observations and find that they can explain the bulk of the observations with this model. Therefore, they can use this model to assess the gas abundance in CO, carbon and oxygen around all debris disc stars. Fig.~\ref{kral:fig7}
shows their predictions for the detectability of neutral carbon with APEX and ALMA. They predict that CI around $\beta$ Pic should be detected with ALMA for the on-going observation (PI: Brandeker) and give predictions for the rest of the sample. They find
that ALMA could revolutionise our understanding of gas around debris discs and predict detections of neutral carbon in at least 30 systems.

\section{Conclusions}

In this proceeding, the reader is given an understanding of the physics at work in debris discs as well as a state-of-the-art compilation of the different modelling methods 
that are used among the debris disc community. It is not our aim to review every single study but rather give an overview of the different modelling possibilities and the most recent works
that have been published in our community.

The new codes coupling dynamics and collisions are a great step forward in terms of modelling debris discs and open a new era where these discs can be modelled self-consistently and taking into account their full complexity.
We also emphasise that there is still room for new simpler approaches such as \citet{2016ApJ...827..125L}, where they use a simple debris disc model with a planet and reproduce a wide variety of disc morphologies that are observed. 
Gas observations in debris disc systems could unveil a totally new independent picture of planetary systems compared to dust observations. This new field is emerging quickly and thanks to the high-resolution power of ALMA could lead
to great results in the close future. For instance, atomic gas is expected to extend down to the central star whilst debris disc are located at tens of au. Observing these gas discs could be a way to probe, for the first time, the hidden inner
parts of planetary systems and could potentially reveal some hidden planets through structures or asymmetries on the gas disc at a few au.  

\begin{acknowledgements}
QK acknowledges support from the European Union through ERC grant number 279973 and wishes to thank the PNP and the ASA for the invitation to give this talk. I also thank A. Bonsor and P. Thebault for interesting discussions.
\end{acknowledgements}



%
\end{document}